\documentclass[11pt,a4paper]{article}

\usepackage[T1]{fontenc}
\usepackage[utf8]{inputenc} 
\usepackage{graphicx}
\usepackage{amsmath,amssymb}
\usepackage{cite}
\usepackage[colorlinks=true,linkcolor=blue,citecolor=blue,urlcolor=blue]{hyperref}
\usepackage{orcidlink}
\usepackage{QCDEX2026}
\usepackage{subfig}

\begin{document}

\noindent\includegraphics[width=0.4\textwidth]{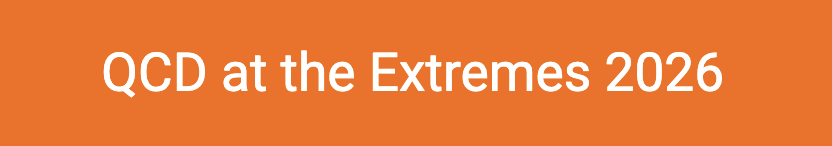}\par
\vspace{0.6em}
\begin{center}
  {\LARGE\bfseries The quantum magic of top quarks\par}
  \vspace{0.8em}
  {\large
    Chris D. White$^{1,*}$  
    \par}
  \vspace{0.5em}
  {\small
    $^{1}$School of Physical and Chemical Sciences, Queen Mary University of London, London, United Kingdom\par}
  \vspace{0.35em}
  {\small $^{*}$Corresponding author: \href{mailto:christopher.white@qmul.ac.uk}{christopher.white@qmul.ac.uk}\par}
  \vspace{0.35em}
  {\small\itshape refereed by the conveners of QCDEX2026\par}
\end{center}
\vspace{0.5em}

\begin{abstract}
In recent years, a growing community of researchers has been applying
ideas from Quantum Information theory to high energy
physics. Entanglement is one such quantity, but there are by now many
others. This contribution will examine the property of “magic” (also
known as “non-stabliserness”) which, roughly speaking, measures
whether quantum systems have a genuine computational advantage over
their classical counterparts. We will review the origin of this
concept, before describing recent applications in collider physics,
including top quark production. 

\end{abstract}


\section{Introduction}
\label{sec:intro}

Quantum Information (QI) theory is by a now a well-established field,
with a number of applications. On the one hand, it underlies the
design and implementation of quantum computers, and the development of
practical algorithms for such devices. On the other hand, QI methods
provide a general language for examining quantum systems, ranging from
low-energy table top experiments, through to high energy particle
physics facilities such as the Large Hadron Collider. A by now
well-studied property at colliders is that of quantum
entanglement\footnote{The study of entanglement -- and its
relationship with other quantum properties such as Bell inequalities
-- is subject to ongoing debate (see
e.g. Refs.~\cite{Abel:1992kz,Abel:2025wwa,Bechtle:2025ugc,Aguilar-Saavedra:2026rsx,ATLAS:2026bre,Low:2025aqq}). No
such controversies affect the results of this contribution, which
assumes the framework of quantum mechanics throughout.}, with a key
example process being that of top quark pair
production~\cite{Afik:2020onf,Afik:2022kwm,Dong:2023xiw,Fabbrichesi:2021npl,Aoude:2022imd,Aguilar-Saavedra:2022uye}. Relevant
experimental analyses include Refs.~\cite{ATLAS:2023fsd,CMS:2024pts},
and an excellent pedagogical review of this area is that of
Ref.~\cite{Barr:2024djo}.

The previous wide interest in entanglement has paved the way, more
recently, for the application of other ideas from QI theory to a high
energy setting. There are a number of motivations. First, there is the
possibility of testing quantum mechanics itself (subject to loopholes
and other caveats). Secondly, there is the hope that QI-motivated
observables may help in disentangling new physics theories from the
current Standard Model of Particle Physics. Thirdly, there is an
increasing body of work arguing that QI methods could explain
structures within our existing theories, or their possible
extensions. More generally, QI theory may give us new ways of thinking
about open problems in high energy physics theory or phenomenology.

In this contribution, we will focus on a particular QI property known
as {\it magic} (more formally, {\it non-stabiliserness}). This was
first introduced in Ref.~\cite{Bravyi:2004isx} in the context of
quantum computing, where it is of crucial importance for
characterising the computational advantage of quantum computers over
their classical counterparts, as well as designing fault-tolerant
algorithms. Thus, in order to formally define magic, we must first
briefly review relevant aspects of quantum computing.

\section{A bit of quantum computing}
\label{sec:QC}

\begin{figure}[htbp]
  \centering
   \includegraphics[width=0.65\textwidth]{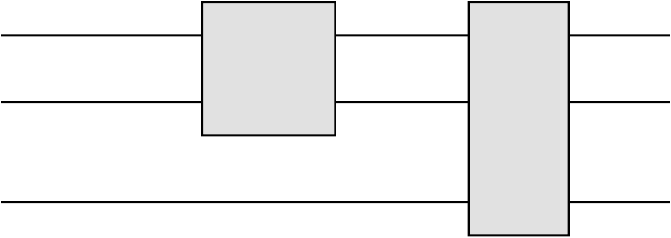}
  \caption{An example quantum circuit diagram, where the incoming
    lines represent the basis state $|00\ldots 0\rangle$, and the
    boxes are quantum gates.}
  \label{fig:circuit}
\end{figure}

Classical computers consist of logical {\it bits} which can take
values 0 or 1. A quantum computer replaces these with {\it
  qubits}\footnote{More generally, quantum computers may contain
individual sates in a Hilbert space with dimension greater than two,
referred to as {\it qudits}.}, whose most general state is a
superposition of two basis states $|0\rangle$ and $|1\rangle$:
\begin{equation}
  |\psi\rangle=\alpha|0\rangle +\beta|1\rangle.
  \label{psidef}
\end{equation}
Here $\alpha$ and $\beta$ are complex coefficients subject to the
normalisation condition
\begin{equation}
  |\alpha|^2+|\beta|^2=1.
  \label{alphabeta}
\end{equation}
A canonical physical example of a qubit is that of a single spin-1/2
particle, such that the states $|0\rangle$ and $1\rangle$ correspond
to the up and down spin states with respect to a chosen direction. A
general quantum computer may contain many qubits, in which case a
suitable choice of basis states is
\begin{equation}
  |\psi_1\psi_2\ldots\psi_n\rangle\equiv
  |\psi_1\rangle\otimes |\psi_2\rangle\otimes\ldots\otimes|\psi_n\rangle.
  \label{multiqubit}
\end{equation}
Quantum computers then take (superpositions of) such states, and
subject them to unitary transformations, where the unitary property
follows from the basic postulates of quantum mechanics. It is
conventional to draw the action of such transformations using quantum
circuit diagrams, an example of which is shown in
figure~\ref{fig:circuit}. Each line in the diagram represents a single
qubit, where the conventional incoming state is $|00\ldots
0\rangle$. Boxes represent unitary transformations, and the two shown
in Fig.~\ref{fig:circuit} act on two and three qubits
respectively. The transformations are then referred to as {\it
  quantum gates}, by analogy with the well-known logic gates of
classical computing. Common gates used for quantum computing have
names like {\it Hadamard}, {\it phase}, {\it CNOT} and {\it Pauli}. We
will not need their precise details for what follows, but it is worth
noting that these particular gates tend to be the easiest ones to make
in current quantum systems.

We are interested in quantum computers because they are expected to
vastly outperform their classical counterparts. Na\"{i}vely, this is
due to the two quantum properties of {\it superposition} and {\it
  entanglement}. The former implies that one may have many
calculations going on simultaneously, and the latter is a known
ingredient in quantum algorithms that are known to be exponentially
faster than classical ones for certain problems. However, it is not
actually true that entanglement by itself guarantees higher computing
power. To see why, one may introduce the notion of a {\it stabiliser
  state}. The starting point is to consider the following set of
operators, defined for any number of qubits:
\begin{equation}
  {\cal P}_N=P_1\otimes P_2\otimes\ldots\otimes P_N,\quad
  P_a\in\{\sigma_1^{(a)},\sigma_2^{(a)},\sigma_3^{(a)},I^{(a)}\}.
  \label{Paulistrings}
\end{equation}
These are called {\it Pauli strings}, and consist of a series of
transformations acting simultaneously on each individual qubit, and
where the operator acting on the $a^{\rm th}$ qubit is either an
identity matrix or a Pauli matrix. There are then $4^N$ such operators
for $N$ qubits. Then, given a state $|\psi\rangle$, its {\it Pauli
  spectrum} consists of the set of expectation values for every Pauli
string:
\begin{equation}
  {\rm spec}(|\psi\rangle)=\{\langle \psi|P|\psi\rangle,\quad P\in{\cal P}_N\}.
  \label{Paulispec}
\end{equation}
Stabiliser states are such that $2^N$ expectation values are $+1$ or
$-1$, with the rest being zero. Non-stabiliser states will have a
mixture of different values in the Pauli spectrum. Despite this very
abstract definition, stabiliser states can be interpreted physically
as those states one makes by using the Hadamard, phase, CNOT and Pauli
gates mentioned earlier. Their importance stems from a result called
the {\it Gottesman-Knill} theorem, which one may paraphrase as
follows: {\it For every quantum computer containing stabiliser states
  only, there is a classical computer that is just as efficient}.

Put more bluntly: there is no computational advantage to be gained by
building an expensive quantum computer, if all it contains are the
very gates that are easiest to make in the lab! Unsurprisingly, this
tells us that quantum computing is a rather difficult endeavour. The
result can also be puzzling when first encountered, as it turns out
that stabiliser states include maximally entangled ones (e.g. the
well-known Bell states for two qubits). A priori, one might think that
the more entanglement one has, the more powerful one's quantum
computer is. The Gottesman-Knill theorem tells us that this is not
true at all, and it follows that something other than entanglement is
needed for computational advantage. It is this resource that has been
called {\it magic} in the QI literature. As a term, it is synonymous
with {\it non-stabliserness} of a quantum state, which is what it
ultimately measures.

Various definitions of magic exist in the literature, motivated by
ease of computation or measurement in particular scenarios. A popular
definition is the set of so-called {\it Stabiliser R\'{e}nyi
  entropies}, first presented in Ref.~\cite{Leone:2021rzd}:
\begin{equation}
  M_q=\frac{1}{1-q}\log_2(\zeta_q),\quad \zeta_q\equiv
  \sum_{P\in{\cal P}_N}\frac{\langle \psi|P|\psi\rangle^{2q}}{2^N}.
\label{Mqdef}
\end{equation}
There is one such quantity for each value of the integer $q\geq 2$,
and we see that each involves the Pauli spectrum (through expectation
values of Pauli strings $P$), as it should given that deviations from
a simple Pauli spectrum constitute non-stabiliserness. One may check
that the magic defined in this way indeed vanishes for stabiliser
states, and it is also additive when multiqubit systems are
combined. For what follows, choosing $q=2$ will suffice, and the
quantity $M_2$ is then usually referred to as the {\it Second
  Stabiliser R\'{e}nyi Entropy (SSRE)}.

\section{The magic of top quarks}
\label{sec:magictop}

\begin{figure}[htbp]
  \centering
   \includegraphics[width=0.4\textwidth]{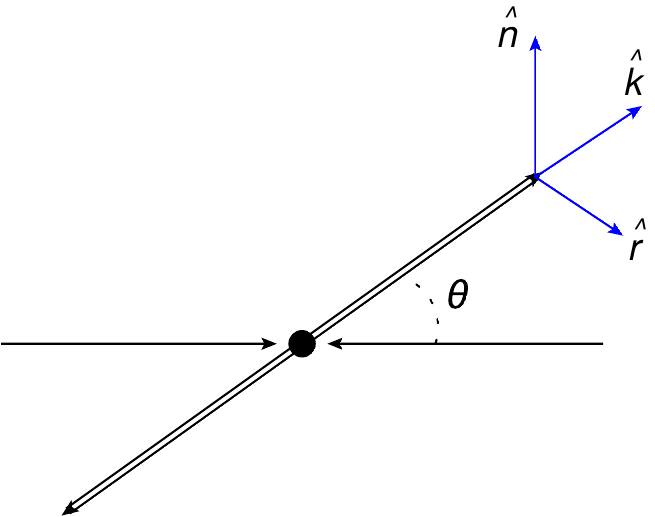}
  \caption{In the helicity basis, the top quark spin directions are
    mapped to spatial directions parallel or orthogonal to the top
    quark itself. The scattering angle $\theta$ is defined with
    respect to the first incoming beam.}
  \label{fig:helicity}
\end{figure}

\begin{figure}
    \centering
    \subfloat[]{\includegraphics[width=0.45\textwidth]{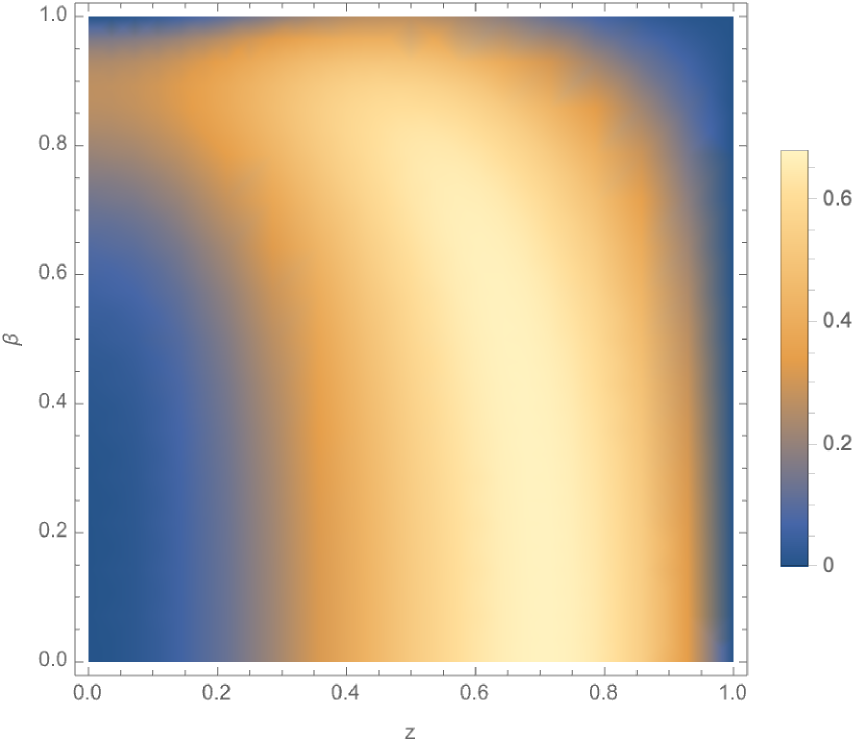} }
    \qquad
    \subfloat[]{\includegraphics[width=0.45\textwidth]{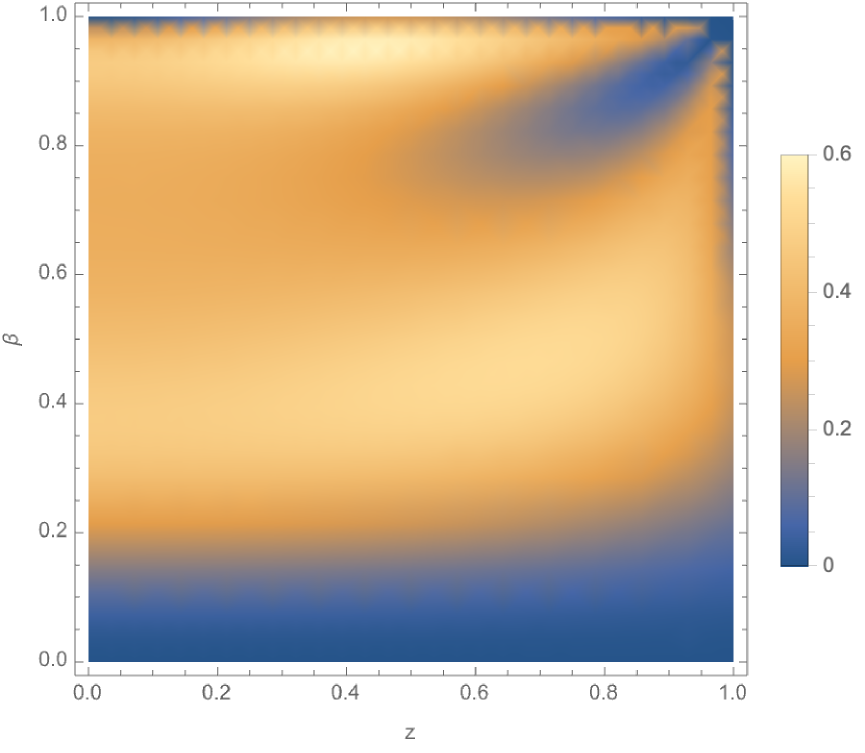} }
    \qquad
    \subfloat[]{\includegraphics[width=0.45\textwidth]{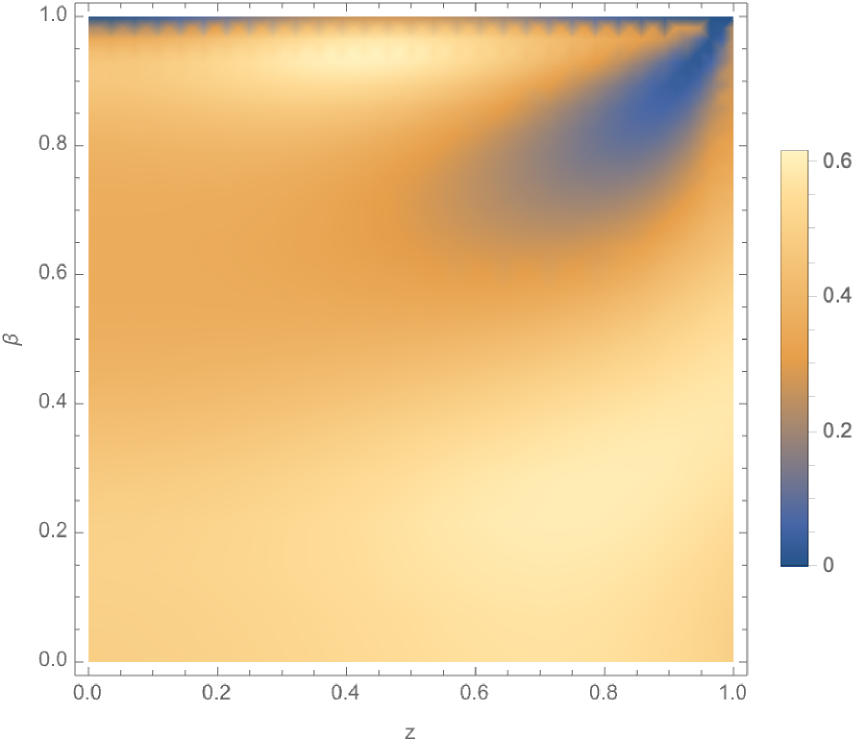} }
    \caption{The magic of a mixed top-antitop final state in: (a) the
      $q\bar{q}$ channel; (b) the $gg$ channel; (c) for initial state
      protons.}
    \label{fig:magic}
\end{figure}

There is by now a well-established and vibrant global community of
researchers interested in many-body quantum magic, with even an entire
annual interdisciplinary conference devoted to the
subject~\footnote{The Many Body Quantum Magic (MBQM) conference was
held in Abu Dhabi, Seattle and Trento in 2024, 2025 and 2026
respectively.}. As noted above, magic plays a key role in generating
computational advantage and fault tolerance of quantum computers. It
has also become increasingly clear that magic is an important tool for
characterising interesting behaviours such as phase transitions and
confinement (see e.g. ref.~\cite{Robin:2026lqp} for a recent
review). People are thus interested in any physical system where magic
can be produced and / or enhanced, so that new insights can be
obtained, which may in turn be portable to other quantum systems. High
energy scattering processes provide one such playground, and the Large
Hadron collider excels in particular at generating final states
containing two qubits (spin-1/2 particles). A key example is that of
top quark pair production, already mentioned above in the context of
entanglement studies. At leading order in the Standard Model, top
quarks are produced form either a quark-antiquark or gluon-gluon
initial state, and the final state will consist of different possible
pure quantum states, weighted by different probabilities as dictated
by the rules of quantum field theory. This is a so-called {\it mixed
  state}, and one must generalise the above measure of magic
accordingly. First, one calculates a (spin) density matrix for the
final state, schematically given by
\begin{equation}
  \rho_{\alpha\beta,\alpha'\beta'}\propto \left\langle
      {\cal A}_{\alpha\beta}\left[
        {\cal A}_{\alpha'\beta'}
        \right]
  \right\rangle,
  \label{rhodef}
\end{equation}
where ${\cal A}_{\alpha\beta}$ is the scattering amplitude with
final-state spin indices $\alpha,\beta$, and the angle brackets denote
averaging / projection over non-spin quantum numbers (see
Refs.~\cite{Afik:2020onf,Afik:2022kwm} for further details). Armed
with the density matrix, one may calculate the mixed-state SSRE via
\begin{equation}
  \tilde{M}_2(\rho)=-\log_2\left(\frac{\sum_{P\in{\cal P}_n}
    {\rm Tr}^4(\rho P)}{\sum_{P\in{\cal P}_n} {\rm Tr}^2(\rho P)}\right).
\label{M2tilde}
\end{equation}
The density matrix itself depends on the basis chosen for identifying
spin directions with physical directions in space. A common choice is
the helicity basis (see e.g. Ref.~\cite{Baumgart:2012ay}), which
aligns the spin direction with the outgoing top quark, as shown in
fig.~\ref{fig:helicity}. One may then express the amount of magic
(SSRE) as a function of the cosine of the scattering angle and the
4-velocity of the top quark (in units of the speed of light). These
are given respectively by
\begin{equation}
  z=\cos\theta,\quad \beta=\sqrt{1-\frac{4m_t^2}{\hat{s}}},
 \label{zbeta}
\end{equation}
where $m_t$ is the top quark mass, and $\hat{s}$ the partonic centre
of mass energy. As explained in more detail in
Ref.~\cite{White:2024nuc}, this now provides all the ingredients
needed to compute the magic of top quark final states at the
LHC. Results are presented in fig.~\ref{fig:magic} at leading order in
the Standard Model. One sees that nature chooses to produce magic top
quarks over the majority of the final state phase space where,
unsurprisingly, the profile of magic is different for different
initial states. For both the $q\bar{q}$ and $gg$ channels, zero magic
is observed in certain kinematic regions, and one may indeed verify
analytically that the corresponding states are
stabiliser~\cite{White:2024nuc}. 


\begin{figure}
    \begin{center}
        \scalebox{0.4}{\includegraphics{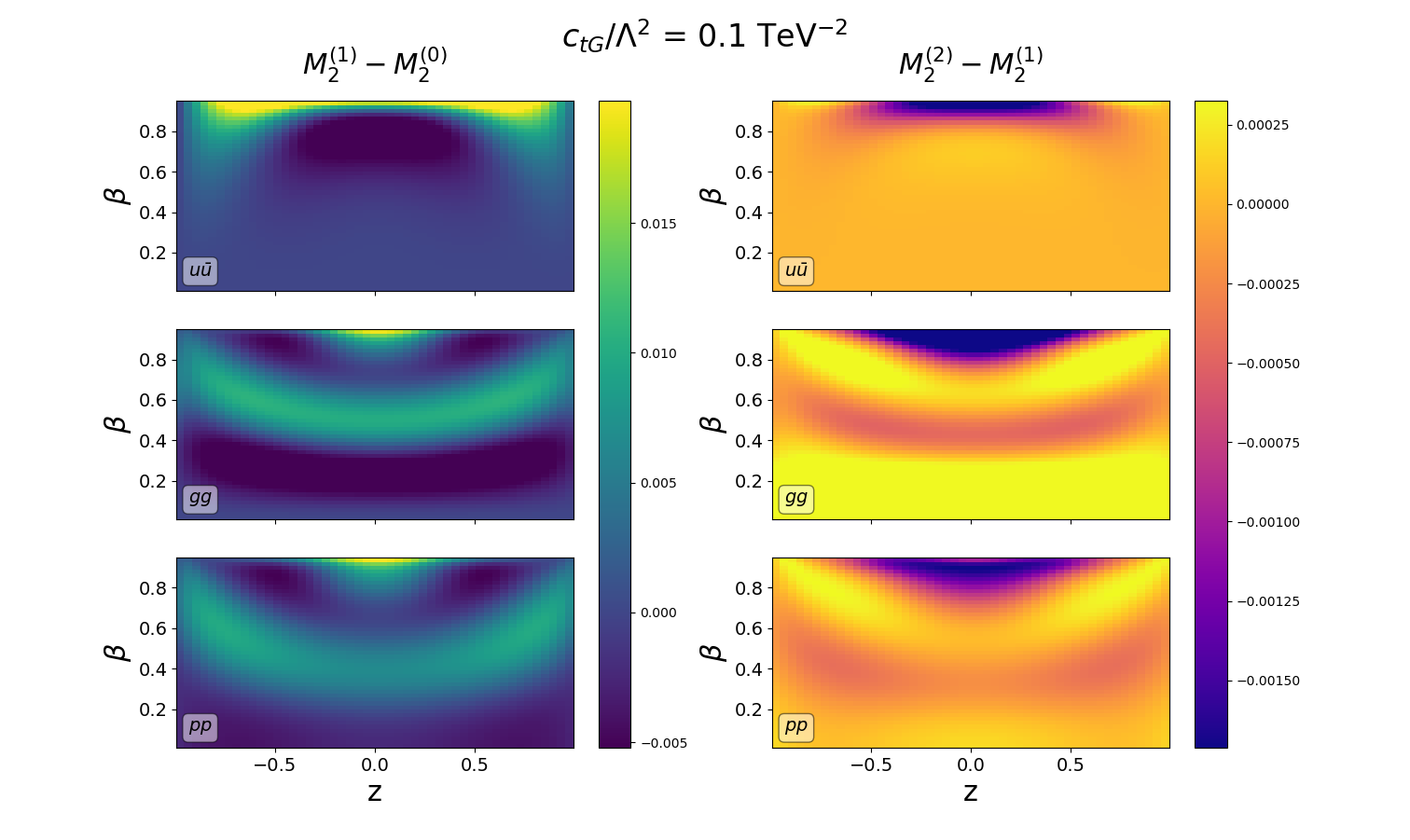}}
        \caption{Two different ways of measuring the difference in
          magic for a new physics theory (corresponding to a single
          additional SMEFT operator added to the Standard
          Model). Figure taken from Ref.~\cite{Aoude:2025jzc}.}
        \label{fig:ctG}
    \end{center}
\end{figure}

Reference~\cite{Fabbrichesi:2025ywl} studied whether magic -- and
other quantum information quantities -- can be used to disentangle new
physics from the Standard Model. This was investigated in the specific
context of top quark production in Ref.~\cite{Aoude:2025jzc}. The
authors used the well-known Standard Model Effective Theory (SMEFT) to
add additional corrections to the Standard Model, and examined how the
magic profile changed when such contributions are included. An example
is shown in figure~\ref{fig:ctG}, showing that the magic profile
indeed changes when new physics is
present. Reference~\cite{Aoude:2025jzc} also looked at whether magic
or alternative QI measures (e.g. so-called trace distance and
fidelity) are more or less sensitive as probes of new physics,
including compared to simply using spin correlation coefficients by
themselves. Depending on the kinematic region or specific new physics
type, any of the examined measures can end up being most sensitive,
corroborating similar conclusions in different scattering processes
from Ref.~\cite{Fabbrichesi:2025ywl}. Such results are also of
interest to the general many body magic community, who often study
magic in systems that can be deformed by varying one or more
parameters. Related studies of magic or similar quantities in a
variety of quantum systems include
Refs.~\cite{Gargalionis:2025iqs,Gargalionis:2026onv,Aoude:2022imd,Aoude:2023hxv,Aoude:2025jzc,Fabbrichesi:2025ywl,Liu:2025qfl,Beverland_2020,PhysRevApplied.19.034052,Leone:2023avk,Qassim2021improvedupperbounds,Leone:2021rzd,Haug:2023ffp,Magic1,PhysRevA.108.042408,Gu:2023qqq,Tirrito:2023fnw,Turkeshi:2023lqu,Leone:2024lfr,Tarabunga:2023ggd,Frau:2024qmf,Lami:2024osd,Robin:2024oqc,Jha:2026ror,Jha:2026eni,Chu:2026euw,Chu:2026yxm}.

In parallel with theoretical studies of magic in scattering processes,
experimental collaborations have also become interested. For example,
Ref.~\cite{CMS:2025cim} presents a differential measurement of magic
in top pair final states by the CMS collaboration, in bins of
invariant mass and scattering angle. The ATLAS experiment has produced
a general document outlining its plan for QI measurements, which
explicitly lists magic as an interesting quantity for further
study~\cite{ATLAS:2026bre}. A more comprehensive study is that of
Ref.~\cite{Afik:2025ejh}, an input to a recent update of the European
strategy for particle physics. Two particular themees that emerge are
the use of collider experiments to probe quantum measurements that can
be difficult to reconstruct in traditional table top experiments; and
the development of better analysis tools for new physics.

\section{Summary and outlook}

Magic (also known as non-stabiliserness) is a well-studied concept in
quantum information theory, due to its crucial importance for
developing efficient and fault-tolerant quantum computers. Recent
years have seen widespread study of magic in various quantum systems,
where it seems to be an efficient marker for novel behaviours
including phase transitions and confinement. In this contribution, we
have examined recent work demonstrating that magic is naturally
produced in top quark final states at the LHC, and can also be a
useful probe of new physics. This itself forms part of a wider
dialogue between the fields of QI theory and particle physics, that
promises to be a fruitful activity in the coming years. High energy
physics offers new playgrounds (both theoretical and experimental) for
QI theorists to sharpen their tools and insights. What is often
invisible to pure particle physics audiences is the high level of
interest for high energy physics scenarios that has arisen amongst the
global community of people working on many body quantum
magic. Likewise, both high energy theorists and experimenters can
benefit from knowing more about the language of quantum information
theory. The coming years are likely to see rapid developments in this
area, and increased interdisciplinary collaboration.

\section*{Acknowledgements}
This research was supported by the UK Science and Technology
Facilities Council (STFC) Consolidated Grant ST/P000754/1 ``String
theory, gauge theory and duality''.

\bibliographystyle{QCDEX2026}
\bibliography{QCDEX2026}

\providecommand{\href}[2]{#2}\begingroup\raggedright\begin{thebibliography}{10}%
\makeatletter
\providecommand{\hrefCMSnoop }[0]{\@secondoftwo}%
\makeatother
\providecommand{\doi}{\texttt{doi:}\begingroup \urlstyle{tt}\Url}

\bibitem{Abel:1992kz}
\hrefCMSnoop {}{S.~A. Abel, M.~Dittmar, and H.~K. Dreiner, ``{Testing locality
  at colliders via Bell's inequality?}'',} \textit{ Phys. Lett. B} \textbf{
  280} (1992) 304--312.

\bibitem{Abel:2025wwa}
\hrefCMSnoop {}{S.~A. Abel, H.~K. Dreiner, R.~Sengupta, and L.~Ubaldi,
  ``{Colliders are not testing locality via Bell's inequality nor providing an
  unconditional proof of entanglement}'',} \textit{ JHEP} \textbf{ 08} (2026)
  067, \href{http://www.arXiv.org/abs/2507.15949}{\texttt{arXiv:2507.15949}}.

\bibitem{Bechtle:2025ugc}
\hrefCMSnoop {}{P.~Bechtle, C.~Breuning, H.~K. Dreiner, and C.~Duhr, ``{A
  critical appraisal of tests of locality and of entanglement versus
  non-entanglement at colliders}'',}
  \href{http://www.arXiv.org/abs/2507.15947}{\texttt{arXiv:2507.15947}}.

\bibitem{Aguilar-Saavedra:2026rsx}
\hrefCMSnoop {}{J.~A. Aguilar-Saavedra, J.~A. Casas, and J.~M. Moreno,
  ``{Understanding Bell locality tests at colliders}'',} \textit{ Eur. Phys. J.
  C} \textbf{ 86} (2026), no.~7, 868,
  \href{http://www.arXiv.org/abs/2603.19389}{\texttt{arXiv:2603.19389}}.

\bibitem{ATLAS:2026bre}
\hrefCMSnoop {}{{ATLAS} Collaboration, ``{Probing Quantum Information
  observables with the ATLAS detector: some general principles}'',}.

\bibitem{Low:2025aqq}
\hrefCMSnoop {}{M.~Low, ``{Addressing local realism through Bell tests at
  colliders}'',} \textit{ Phys. Rev. D} \textbf{ 112} (2025), no.~9, 096008,
  \href{http://www.arXiv.org/abs/2508.10979}{\texttt{arXiv:2508.10979}}.

\bibitem{Afik:2020onf}
\hrefCMSnoop {}{Y.~Afik and J.~R.~M. de~Nova, ``{Entanglement and quantum
  tomography with top quarks at the LHC}'',} \textit{ Eur. Phys. J. Plus}
  \textbf{ 136} (2021), no.~9, 907,
  \href{http://www.arXiv.org/abs/2003.02280}{\texttt{arXiv:2003.02280}}.

\bibitem{Afik:2022kwm}
\hrefCMSnoop {}{Y.~Afik and J.~R.~M. de~Nova, ``{Quantum information with top
  quarks in QCD}'',} \textit{ Quantum} \textbf{ 6} (2022) 820,
  \href{http://www.arXiv.org/abs/2203.05582}{\texttt{arXiv:2203.05582}}.

\bibitem{Dong:2023xiw}
\hrefCMSnoop {}{Z.~Dong, D.~Gon{\c{c}}alves, K.~Kong, and A.~Navarro,
  ``{Entanglement and Bell inequalities with boosted t$\bar{t}$}'',} \textit{
  Phys. Rev. D} \textbf{ 109} (2024), no.~11, 115023,
  \href{http://www.arXiv.org/abs/2305.07075}{\texttt{arXiv:2305.07075}}.

\bibitem{Fabbrichesi:2021npl}
\hrefCMSnoop {}{M.~Fabbrichesi, R.~Floreanini, and G.~Panizzo, ``{Testing Bell
  Inequalities at the LHC with Top-Quark Pairs}'',} \textit{ Phys. Rev. Lett.}
  \textbf{ 127} (2021), no.~16, 161801,
  \href{http://www.arXiv.org/abs/2102.11883}{\texttt{arXiv:2102.11883}}.

\bibitem{Aoude:2022imd}
\hrefCMSnoop {}{R.~Aoude, E.~Madge, F.~Maltoni, and L.~Mantani, ``{Quantum
  SMEFT tomography: Top quark pair production at the LHC}'',} \textit{ Phys.
  Rev. D} \textbf{ 106} (2022), no.~5, 055007,
  \href{http://www.arXiv.org/abs/2203.05619}{\texttt{arXiv:2203.05619}}.

\bibitem{Aguilar-Saavedra:2022uye}
\hrefCMSnoop {}{J.~A. Aguilar-Saavedra and J.~A. Casas, ``{Improved tests of
  entanglement and Bell inequalities with LHC tops}'',} \textit{ Eur. Phys. J.
  C} \textbf{ 82} (2022), no.~8, 666,
  \href{http://www.arXiv.org/abs/2205.00542}{\texttt{arXiv:2205.00542}}.

\bibitem{ATLAS:2023fsd}
\hrefCMSnoop {}{{ATLAS} Collaboration, ``{Observation of quantum entanglement
  with top quarks at the ATLAS detector}'',} \textit{ Nature} \textbf{ 633}
  (2024), no.~8030, 542--547,
  \href{http://www.arXiv.org/abs/2311.07288}{\texttt{arXiv:2311.07288}}.

\bibitem{CMS:2024pts}
\hrefCMSnoop {}{{CMS} Collaboration, ``{Observation of quantum entanglement in
  top quark pair production in proton{\textendash}proton collisions at
  $\sqrt{s} = 13$ TeV}'',} \textit{ Rept. Prog. Phys.} \textbf{ 87} (2024),
  no.~11, 117801,
  \href{http://www.arXiv.org/abs/2406.03976}{\texttt{arXiv:2406.03976}}.

\bibitem{Barr:2024djo}
A.~J. Barr\hrefCMSnoop {}{ {et~al.}, ``{Quantum entanglement and Bell
  inequality violation at colliders}'',} \textit{ Prog. Part. Nucl. Phys.}
  \textbf{ 139} (2024) 104134,
  \href{http://www.arXiv.org/abs/2402.07972}{\texttt{arXiv:2402.07972}}.

\bibitem{Bravyi:2004isx}
\hrefCMSnoop {}{S.~Bravyi and A.~Kitaev, ``{Universal quantum computation with
  ideal Clifford gates and noisy ancillas}'',} \textit{ Phys. Rev. A} \textbf{
  71} (2005), no.~2, 022316,
  \href{http://www.arXiv.org/abs/quant-ph/0403025}{\texttt{arXiv:quant-ph/0403025}}.

\bibitem{Leone:2021rzd}
\hrefCMSnoop {}{L.~Leone, S.~F.~E. Oliviero, and A.~Hamma, ``{Stabilizer
  R{\'e}nyi Entropy}'',} \textit{ Phys. Rev. Lett.} \textbf{ 128} (2022),
  no.~5, 050402,
  \href{http://www.arXiv.org/abs/2106.12587}{\texttt{arXiv:2106.12587}}.

\bibitem{Robin:2026lqp}
\hrefCMSnoop {}{C.~E.~P. Robin and M.~J. Savage, ``{Quantum Complexity and New
  Directions in Nuclear Physics and High-Energy Physics Phenomenology}'',}
  \href{http://www.arXiv.org/abs/2604.26376}{\texttt{arXiv:2604.26376}}.

\bibitem{Baumgart:2012ay}
\hrefCMSnoop {}{M.~Baumgart and B.~Tweedie, ``{A New Twist on Top Quark Spin
  Correlations}'',} \textit{ JHEP} \textbf{ 03} (2013) 117,
  \href{http://www.arXiv.org/abs/1212.4888}{\texttt{arXiv:1212.4888}}.

\bibitem{White:2024nuc}
\hrefCMSnoop {}{C.~D. White and M.~J. White, ``{Magic states of top quarks}'',}
  \textit{ Phys. Rev. D} \textbf{ 110} (2024), no.~11, 116016,
  \href{http://www.arXiv.org/abs/2406.07321}{\texttt{arXiv:2406.07321}}.

\bibitem{Fabbrichesi:2025ywl}
\hrefCMSnoop {}{M.~Fabbrichesi, M.~Low, and L.~Marzola, ``{Trace distance
  between density matrices: A nifty tool in new-physics searches}'',} \textit{
  Phys. Rev. D} \textbf{ 112} (2025), no.~1, 013003,
  \href{http://www.arXiv.org/abs/2501.03311}{\texttt{arXiv:2501.03311}}.

\bibitem{Aoude:2025jzc}
\hrefCMSnoop {}{R.~Aoude, H.~Banks, C.~D. White, and M.~J. White, ``{Probing
  new physics in the top sector using quantum information}'',} \textit{ Phys.
  Rev. D} \textbf{ 113} (2026), no.~11, 115066,
  \href{http://www.arXiv.org/abs/2505.12522}{\texttt{arXiv:2505.12522}}.

\bibitem{Gargalionis:2025iqs}
J.~Gargalionis\hrefCMSnoop {}{ {et~al.}, ``{Spin versus nonstabilizerness in
  gluon and graviton scattering}'',} \textit{ Phys. Rev. D} \textbf{ 113}
  (2026), no.~1, 016007,
  \href{http://www.arXiv.org/abs/2508.14967}{\texttt{arXiv:2508.14967}}.

\bibitem{Gargalionis:2026onv}
J.~Gargalionis\hrefCMSnoop {}{ {et~al.}, ``{Non-local nonstabiliserness in
  Gluon and Graviton Scattering}'',}
  \href{http://www.arXiv.org/abs/2603.04148}{\texttt{arXiv:2603.04148}}.

\bibitem{Aoude:2023hxv}
\hrefCMSnoop {}{R.~Aoude, E.~Madge, F.~Maltoni, and L.~Mantani, ``{Probing new
  physics through entanglement in diboson production}'',} \textit{ JHEP}
  \textbf{ 12} (2023) 017,
  \href{http://www.arXiv.org/abs/2307.09675}{\texttt{arXiv:2307.09675}}.

\bibitem{Liu:2025qfl}
\hrefCMSnoop {}{Q.~Liu, I.~Low, and Z.~Yin, ``{Quantum magic in quantum
  electrodynamics}'',} \textit{ Phys. Rev. D} \textbf{ 113} (2026), no.~5,
  056001,
  \href{http://www.arXiv.org/abs/2503.03098}{\texttt{arXiv:2503.03098}}.

\bibitem{Beverland_2020}
\hrefCMSnoop {}{M.~Beverland, E.~Campbell, M.~Howard, and V.~Kliuchnikov,
  ``Lower bounds on the non-Clifford resources for quantum computations'',}
  \textit{ Quantum Science and Technology} \textbf{ 5} (may, 2020) 035009.

\bibitem{PhysRevApplied.19.034052}
\hrefCMSnoop {}{J.~Jiang and X.~Wang, ``Lower Bound for the $T$ Count Via
  Unitary Stabilizer Nullity'',} \textit{ Phys. Rev. Appl.} \textbf{ 19} (Mar,
  2023) 034052.

\bibitem{Leone:2023avk}
\hrefCMSnoop {}{L.~Leone, S.~F.~E. Oliviero, and A.~Hamma, ``{Learning t-doped
  stabilizer states}'',} \textit{ Quantum} \textbf{ 8} (2024) 1361,
  \href{http://www.arXiv.org/abs/2305.15398}{\texttt{arXiv:2305.15398}}.

\bibitem{Qassim2021improvedupperbounds}
\hrefCMSnoop {}{H.~Qassim, H.~Pashayan, and D.~Gosset, ``Improved upper bounds
  on the stabilizer rank of magic states'',} \textit{ {Quantum}} \textbf{ 5}
  (December, 2021) 606.

\bibitem{Haug:2023ffp}
\hrefCMSnoop {}{T.~Haug, S.~Lee, and M.~S. Kim, ``{Efficient stabilizer
  entropies for quantum computers}'',}
  \href{http://www.arXiv.org/abs/2305.19152}{\texttt{arXiv:2305.19152}}.

\bibitem{Magic1}
\hrefCMSnoop {}{S.~F.~E. Oliviero, L.~Leone, A.~Hamma, and S.~Lloyd,
  ``Measuring magic on a quantum processor'',} \textit{ npj Quantum
  Information} \textbf{ 8} (2022), no.~1, 148.

\bibitem{PhysRevA.108.042408}
\hrefCMSnoop {}{X.~Turkeshi, M.~Schir\`o, and P.~Sierant, ``Measuring
  nonstabilizerness via multifractal flatness'',} \textit{ Phys. Rev. A}
  \textbf{ 108} (Oct, 2023) 042408.

\bibitem{Gu:2023qqq}
A.~Gu\hrefCMSnoop {}{ {et~al.}, ``{Pseudomagic Quantum States}'',} \textit{
  Phys. Rev. Lett.} \textbf{ 132} (2024), no.~21, 210602,
  \href{http://www.arXiv.org/abs/2308.16228}{\texttt{arXiv:2308.16228}}.

\bibitem{Tirrito:2023fnw}
E.~Tirrito\hrefCMSnoop {}{ {et~al.}, ``{Quantifying nonstabilizerness through
  entanglement spectrum flatness}'',} \textit{ Phys. Rev. A} \textbf{ 109}
  (2024), no.~4, L040401,
  \href{http://www.arXiv.org/abs/2304.01175}{\texttt{arXiv:2304.01175}}.

\bibitem{Turkeshi:2023lqu}
\hrefCMSnoop {}{X.~Turkeshi, A.~Dymarsky, and P.~Sierant, ``{Pauli Spectrum and
  Magic of Typical Quantum Many-Body States}'',}
  \href{http://www.arXiv.org/abs/2312.11631}{\texttt{arXiv:2312.11631}}.

\bibitem{Leone:2024lfr}
\hrefCMSnoop {}{L.~Leone and L.~Bittel, ``{Stabilizer entropies are monotones
  for magic-state resource theory}'',} \textit{ Phys. Rev. A} \textbf{ 110}
  (2024), no.~4, L040403,
  \href{http://www.arXiv.org/abs/2404.11652}{\texttt{arXiv:2404.11652}}.

\bibitem{Tarabunga:2023ggd}
\hrefCMSnoop {}{P.~S. Tarabunga, E.~Tirrito, T.~Chanda, and M.~Dalmonte,
  ``{Many-Body Magic Via Pauli-Markov Chains{\textemdash}From Criticality to
  Gauge Theories}'',} \textit{ PRX Quantum} \textbf{ 4} (2023), no.~4, 040317,
  \href{http://www.arXiv.org/abs/2305.18541}{\texttt{arXiv:2305.18541}}.

\bibitem{Frau:2024qmf}
M.~Frau\hrefCMSnoop {}{ {et~al.}, ``{Nonstabilizerness versus entanglement in
  matrix product states}'',} \textit{ Phys. Rev. B} \textbf{ 110} (2024),
  no.~4, 045101,
  \href{http://www.arXiv.org/abs/2404.18768}{\texttt{arXiv:2404.18768}}.

\bibitem{Lami:2024osd}
\hrefCMSnoop {}{G.~Lami, T.~Haug, and J.~De~Nardis, ``{Quantum State Designs
  with Clifford-Enhanced Matrix Product States}'',} \textit{ PRX Quantum}
  \textbf{ 6} (2025), no.~1, 010345,
  \href{http://www.arXiv.org/abs/2404.18751}{\texttt{arXiv:2404.18751}}.

\bibitem{Robin:2024oqc}
\hrefCMSnoop {}{C.~E.~P. Robin and M.~J. Savage, ``{Quantum complexity
  fluctuations from nuclear and hypernuclear forces}'',} \textit{ Phys. Rev. C}
  \textbf{ 112} (2025), no.~4, 044004,
  \href{http://www.arXiv.org/abs/2405.10268}{\texttt{arXiv:2405.10268}}.

\bibitem{Jha:2026ror}
R.~G. Jha\hrefCMSnoop {}{ {et~al.}, ``{Magic and entanglement in
  1+1-dimensional SU(2) lattice gauge theory}'',}
  \href{http://www.arXiv.org/abs/2606.09971}{\texttt{arXiv:2606.09971}}.

\bibitem{Jha:2026eni}
R.~G. Jha\hrefCMSnoop {}{ {et~al.}, ``{Non-stabilizerness and entanglement in
  $(2+1)$-dimensional SU(2) lattice gauge theory using tensor networks}'',}
  \href{http://www.arXiv.org/abs/2609.28634}{\texttt{arXiv:2609.28634}}.

\bibitem{Chu:2026euw}
\hrefCMSnoop {}{J.~Chu and S.~Kharel, ``{Stabilizer R{\'e}nyi Entropy Without a
  Phase Reference}'',}
  \href{http://www.arXiv.org/abs/2610.18955}{\texttt{arXiv:2610.18955}}.

\bibitem{Chu:2026yxm}
\hrefCMSnoop {}{J.~Chu and S.~Kharel, ``{Magic without a phase:
  phase-independent stabilizer R{\'e}nyi entropy in gluon scattering}'',}
  \href{http://www.arXiv.org/abs/2607.13134}{\texttt{arXiv:2607.13134}}.

\bibitem{CMS:2025cim}
\hrefCMSnoop {}{{CMS} Collaboration, ``{Observation of magic states of top
  quark pairs produced in proton-proton collisions at
  $\sqrt{s}=13~\mathrm{TeV}$}'',}.

\bibitem{Afik:2025ejh}
\hrefCMSnoop {}{Y.~Afik {et~al.}, ``{Quantum information meets high-energy
  physics: input to the update of the European strategy for particle
  physics}'',} \textit{ Eur. Phys. J. Plus} \textbf{ 140} (2025), no.~9, 855,
  \href{http://www.arXiv.org/abs/2504.00086}{\texttt{arXiv:2504.00086}}.

\end{thebibliography}\endgroup

\end{document}